\documentclass[%
 aps,pra,reprint,
 superscriptaddress,
 amsmath,amssymb,floatfix
]{revtex4-1}

\usepackage{graphicx}
\graphicspath{{Figure/}}
\usepackage{soul,color}
\usepackage{mathtools}
\usepackage{xfrac}
\usepackage[sort&compress]{natbib}
\usepackage{hyperref}
\hypersetup{pdfborder={0 0 0},allbordercolors={0 0 0},colorlinks=True,allcolors=blue}

\begin{document}

\title{Searching for Materials with High Refractive Index and Wide Band Gap: A First-Principles High-Throughput Study}

\author{Francesco Naccarato}
\affiliation{Institute of Condensed Matter and Nanosciences, Universit\'e Catholique de Louvain, 8 Chemin des \'etoiles, 1348 Louvain-la-Neuve, Belgium}
\affiliation{Physics and Materials Science Research Unit, University of Luxembourg, 162a avenue de la Fa\"iencerie, L-1511 Luxembourg, Luxembourg}
\affiliation{European Theoretical Spectroscopy Facility (ETSF)}

\author{Francesco Ricci}%
\affiliation{Institute of Condensed Matter and Nanosciences, Universit\'e Catholique de Louvain, 8 Chemin des \'etoiles, 1348 Louvain-la-Neuve, Belgium}

\author{Jin Suntivich}
\affiliation{Department of Materials Science and Engineering, Cornell University, Ithaca, New York 14853, USA}
\affiliation{Kavli Institute at Cornell for Nanoscale Science, Cornell University, Ithaca, New York 14853, USA}

\author{Geoffroy Hautier}%
\affiliation{Institute of Condensed Matter and Nanosciences, Universit\'e Catholique de Louvain, 8 Chemin des \'etoiles, 1348 Louvain-la-Neuve, Belgium}

\author{Ludger Wirtz}
\affiliation{Physics and Materials Science Research Unit, University of Luxembourg, 162a avenue de la Fa\"iencerie, L-1511 Luxembourg, Luxembourg}
\affiliation{European Theoretical Spectroscopy Facility (ETSF)}

\author{Gian-Marco Rignanese}%
\affiliation{Institute of Condensed Matter and Nanosciences, Universit\'e Catholique de Louvain, 8 Chemin des \'etoiles, 1348 Louvain-la-Neuve, Belgium}
\affiliation{European Theoretical Spectroscopy Facility (ETSF)}

\date{\today}

\begin{abstract}

Materials combining both a high refractive index and a wide band gap are of great interest for optoelectronic and sensor applications. 
However, these two properties are typically described by an inverse correlation with high refractive index appearing in small gap materials and vice-versa.
Here, we conduct a first-principles high-throughput study on more than 4000 semiconductors (with a special focus on oxides).
Our data confirm the general inverse trend between refractive index and band gap but interesting outliers are also identified.
The data are then analyzed through a simple model involving two main descriptors: the average optical gap and the effective frequency.
The former can be determined directly from the electronic structure of the compounds, but the latter cannot.
This calls for further analysis in order to obtain a predictive model.
Nonetheless, it turns out that the negative effect of a large band gap on the refractive index can counterbalanced in two ways: (i) by limiting the difference between the direct band gap and the average optical gap which can be realized by a narrow distribution in energy of the optical transitions and (ii) by increasing the effective frequency which can be achieved through either a high number of transitions from the top of the valence band to the bottom of the conduction or a high average probability for these transitions.

Focusing on oxides, we use our data to investigate how the chemistry influences this inverse relationship and rationalize why certain classes of materials would perform better.
Our findings can be used to search for new compounds in many optical applications both in the linear and non-linear regime (waveguides, optical modulators, laser, frequency converter, etc.).

\end{abstract}

\maketitle

\section{Introduction}
 
Light-matter interaction is at the core of various technologies (e.g. lasers, liquid-crystal displays, light-emitting diodes, ...) with applications in many sectors (telecommunications, medicine, energy, transistors, microelectronics, ...)~\cite{Rondinelli2015}.
Improvement and further development of these technologies requires a thorough comprehension of the underlying physical processes and how optical properties are linked to the electronic structure.
Hence, the study of the optical properties of materials has always generated considerable interest and curiosity in the scientific community (see Ref.~\citenum{Odom2015} for a compilation of articles about recent developments).
In particular, high refractive index materials are required to improve the performance of optoelectronic devices such as waveguide-based optical circuits, optical interference filters and mirrors, optical sensors, as well as solar cells (e.g. as anti-reflection coatings).
Furthermore, according to the empirical Miller rule, high refractive index materials also potentially show high response in the non linear regime~\cite{Miller1964}.
In addition to having a high refractive index, the materials used in those optoelectronic devices are often required to have a wide band gap.
This guarantees transparency over the visible spectral range and makes it possible for the devices to operate at higher temperatures and to switch at larger voltages~\cite{Kirschman1999}.
As a result, there is a strong push towards developing materials with high refractive index and wide band gap.
However, while there is an abundance of systems with a wide band gap ($E_g \geq 6$~eV), or with a high refractive index ($n \geq 2$), there are unfortunately few materials which satisfy both requirements at the same time.
The main difficulty in devising such compounds is due to the known inverse relationship between refractive index and band gap (see, e.g., Ref.~\citenum{Tripathy2015} for a review of the different empirical or semi-empirical relations that have been proposed).

First-principles calculations have proven to be a very powerful tool to explore the electronic and optical properties of materials.
Density Functional Theory~(DFT)~\cite{Hohenberg1964,Kohn1965} provides a good description of the electronic structure, apart from a systematic underestimation of the band gap with respect to experiments.
Density Functional Perturbation Theory (DFPT)~\cite{Baroni1987, Gonze1995} is widely used to predict the linear response (and related physical quantities) of periodic systems when they are submitted to an external perturbation.
For instance, when considering the effect of a homogeneous electric field, DFPT allows one to compute the macroscopic dielectric function in the static limit ($\omega=0~eV$).
The success of DFT and DFPT stems from their reliability and low computational cost.
As a result, first-principles calculations have recently been combined with a high-throughput~(HT) approach~\cite{Hautier2012,Curtarolo2013} targeting the discovery of new materials.
Indeed, the combination of these two methods enables the creation of large databases of materials properties that would be prohibitive (in time and cost) for experimental measurements. 
By screening those databases, new materials, targeting specific applications, can be identified.
Successful discoveries (i.e., prediction confirmed in the lab) include materials for batteries, hydrogen production and storage,
thermoelectrics and photovoltaics (see, e.g., Refs.~\citenum{Curtarolo2013} and~\citenum{Jain2016}).
Databases can also be analyzed using data mining techniques, aiming at identifying trends that can give a further insight into the
comprehension of the materials properties, or even make predictions for unknown compounds through Machine Learning (see, for example, Refs.~\citenum{Morgan2004} and~\citenum{Hansen2013}). 

In this paper, we investigate the relationship between the refractive index and the band gap using a first-principles HT approach relying on DFT and DFPT.
Our aim is to provide a statistical, ``data driven'', analysis based on a large set of 4040 semiconductors.
Calculated data confirm the global inverse trend between those two properties, as recently discussed in similar works~\cite{Yim2015, Petousis2016, Petousis2017}.
However, there is also a wide spread of the data around this general tendency, pointing out some outliers with both relatively high refractive index and wide band gap among which well-known materials (TiO$_2$, LiNbO$_3$, ...), already widely used for optical applications, and other materials, not yet considered for such applications (Ti$_3$PbO$_7$, LiSi$_2$N$_3$, BeS, ...).
By mapping all the compounds onto a two-state system, a simple model is derived some descriptors of which can be accessed from the electronic structure.
The density of states (DOS) at the valence and conduction band edges as well as the effective masses of those bands are found to play a critical role for achieving a high refractive index and a wide band gap simultaneously.
Indeed, the availability of a large number of weakly dispersive states for optical transitions can partly counterbalance the inverse relationship between the refractive index and the band gap.
Based on these considerations, we focus on the 3375 oxides present in the data set.
We examine these materials in terms of their chemistry and pinpoint the most interesting ones.

\section{High-Throughput computation} \label{HTMethods}

Our database is built as follows.
We start from the relaxed structures available in the Materials Project (MP) repository~\cite{Jain2013}.
Their thermodynamical stability can be assessed by the energy above hull $E_\textrm{hull}$~\cite{Ong2008,Chen2012}: for a stable compound, $E_\textrm{hull}=0$~meV/atom, and the stability decreases as $E_\textrm{hull}$ increases.
Here, we extract the materials with $E_\textrm{hull}\leq25$~meV/atom \cite{Hautier2012a}.
We also include a few exceptions (with $E_\textrm{hull}>25$~meV/atom) already investigated previously in the literature for technological applications.
The 4040 selected materials cover a broad range of chemistries (oxides, fluorides, sulfides, ...) with various compositions (binaries, ternaries, ...).
However, a significant fraction of those (3375 out of 4040) are oxides, since they show important applications in many sectors (semiconductor industry, catalysts, ...) with an exceptionally broad range of electronic properties (see Ref.~\citenum{Cox2010}).

For all those structures, the static part of the refractive index $n_s$ is computed in the framework of DFPT.
All the calculations are performed with the VASP software package~\cite{Kresse1996}, adopting the projector augmented wave (PAW) method~\cite{Gajdos2006}, and using the Generalized Gradient Approximation (GGA) for the exchange-correlation functional as parameterized by Perdew, Burke, and Ernzerhoff (PBE)~\cite{Perdew1996}.
When dealing with oxides including elements with partially occupied $d$ electrons (such as V, Cr, Mn, Fe, Co, Ni, or Mo) a Hubbard-like Coulomb $U$ term is added to the GGA (GGA+$U$)~\cite{Dudarev1998} to correct the spurious GGA self-interaction energies, adopting the $U$ values advised by the MP~\cite{Jain2011a}.
The electronic properties are calculated from the band structures available in the MP~\cite{Ricci2017}.
For the band gap, we focus on the direct band gap, $E_g^d$, since optical processes are related to vertical transitions.
It is worth pointing out that DFT is known to underestimate the band gap up to the 50\% with
respect to experiments (see for example Refs.~\citenum{Agapito2015, Chan2010}), while
a tendency to overestimate $n_s$ is to be expected. Further details on the validation of a similar
workflow and on the error of the refractive index computed via DFPT can be found in
Ref.~\citenum{Petousis2016}.

Finally, the results (dielectric function, refractive index, space groups, etc.) are stored using the
MongoDB database engine.

\section{Results and Discussion}
\label{Results}

\subsection{Global trend}

\begin{figure}[htbp]
\includegraphics{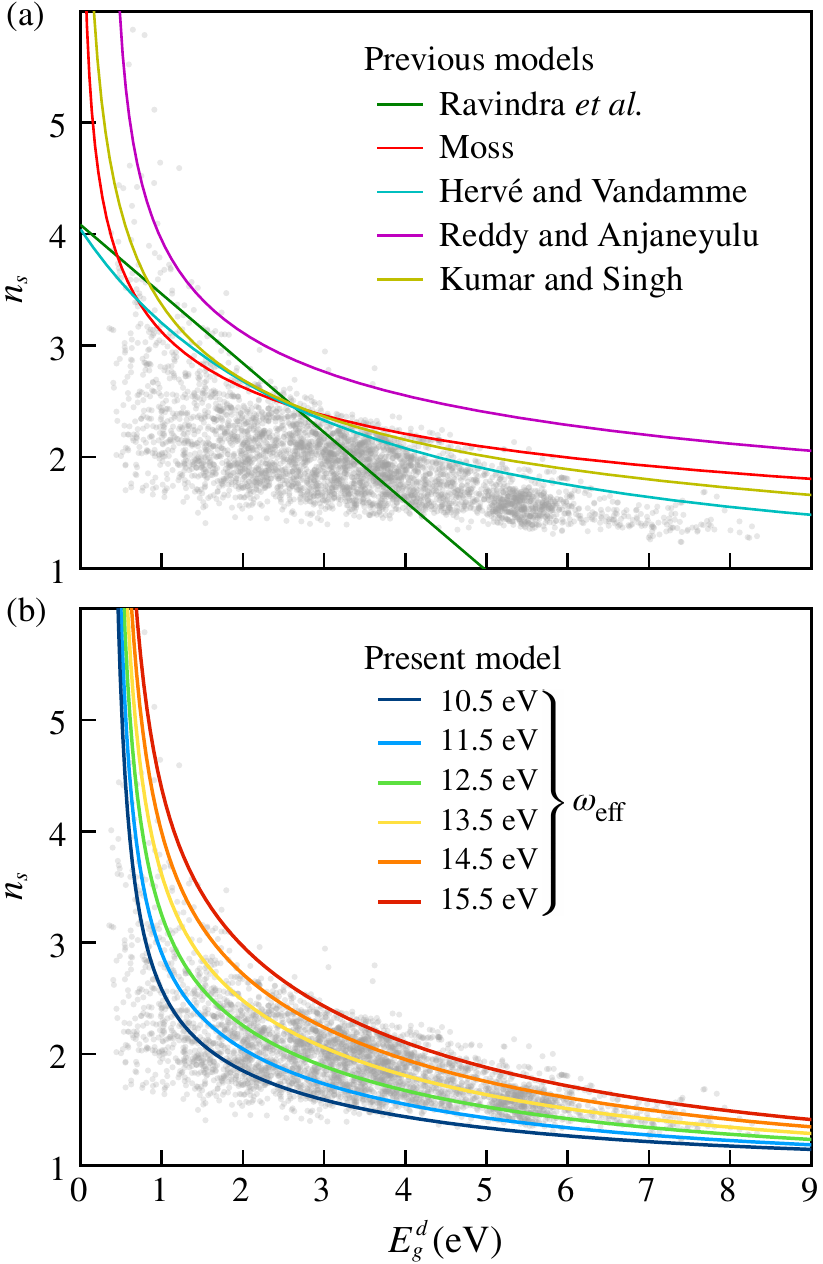}
\caption{Comparison of the calculated data points (refractive index $n_s$ vs. band gap $E_g^d$) with (a) various well-known empirical and semi-empirical models \cite{Ravindra,Moss,Herve1994,Reddy,Kumar2010} and (b) the model described by Eq.~(\ref{eq:ns_vs_eg}).
The data points for the 4040 materials considered here are represented by blue circles, while the models are indicated by solid lines.
In panel (b), different values of the parameter $\omega_{\textrm{eff}}$ have been considered, accounting for the spread in the data points.
}
\label{fig:many_models}
\end{figure}

Various models have been proposed in the literature to describe the inverse relationship between the
refractive index and the band gap.
In Fig.~\ref{fig:many_models}(a), the models proposed by Ravindra \textit{et al.}~\cite{Ravindra} (green line):
\begin{align*}
n_s=4.084-0.62 E_g^d,
\end{align*}
Moss~\cite{Moss} (red line):
\begin{align*}
n_s=\left(\frac{95}{E_g^d}\right)^{1/4},
\end{align*}
Herv{\'e} and Vandamme~\cite{Herve1994} (cyan line):
\begin{align*}
n_s=\sqrt{1+\left(\frac{13.6}{E_g^d+3.47}\right)^{2}},
\end{align*}
Reddy and Anjaneyulu~\cite{Reddy} (magenta line):
\begin{align*}
n_s=\left(\frac{154}{E_g^d-0.365}\right)^{1/4},
\end{align*}
and Kumar and Singh~\cite{Kumar2010} (yellow line):
\begin{align*}
n_s=3.3668\left(E_g^d\right)^{-0.32234}
\end{align*}
are superimposed on our calculated data.
A detailed discussion of these models can be found in Ref.~\citenum{Tripathy2015}.
It is clear that none of them follow closely the trend of the data
(their mean absolute errors (MAE) range from 0.42 to 0.91) nor do they account for the wide spread of the points.
It should however be mentioned that all these models were built up using a small set of experimental data points ($\leq100$).
The model resulting from the present study is reported in Fig.~\ref{fig:many_models}(b).
It captures the trend better than all previous models (MAE=0.33 considering $\omega_{\textrm{eff}}=12.10$ eV, calculated by fitting Eq.(\ref{eq:ns_vs_wg}) in the last square sense) and the spread in the data can be accounted for through the parameter $\omega_\textrm{eff}$, which will be defined below. 

\begin{figure}[htbp]
\includegraphics{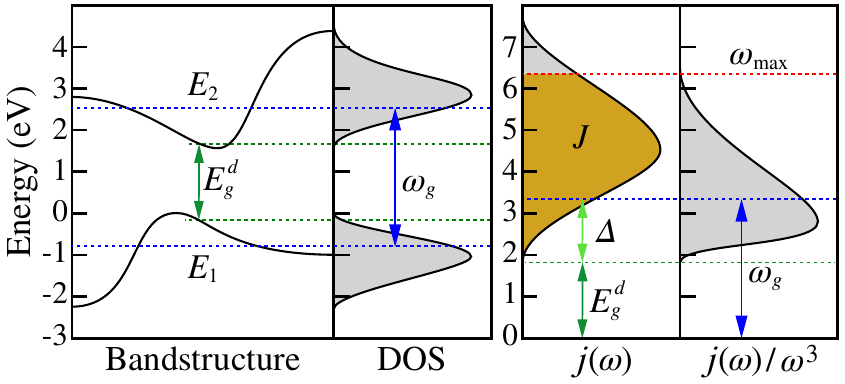}
\caption{
(Left panel) Schematic illustration of the mapping procedure from the electronic structure [band structure and DOS in solid black lines] is replaced by a two-state system ($E_1$, $E_2$ in dashed blue lines).
(Right panel) Optical functions $j(\omega)$, the JDOS, and $j(\omega)/\omega^3$. The direct band gap $E_g^d$ and the average optical gap $\omega_g$ are indicated by green and blue dotted lines, respectively.
The difference $\Delta$ between $\omega_g$ and $E_g^d$ is also reported in light green.
The optical function $j(\omega)/\omega^3$ is used to determine the upper frequency limit $\omega_\textrm{max}$ for the optical absorption processes, as indicated by the red dotted line.
The integral of $j(\omega)$ up to $\omega_\textrm{max}$ leads to the value of $J$,
the degeneracy factor of the transitions between the two states.
}
\label{fig:mapping}
\end{figure}

The inverse relationship between the refractive index and the band gap
is also evident from the following equation: 
\begin{align}\label{eq:epsilon1}
n^2(\omega) & = \varepsilon_1(\omega)
\nonumber \\
& = 1+8\pi \sum_{v,c} 
\int_\mathrm{BZ} & & \frac{2d\textbf{k}}{(2\pi)^3} \frac{\mid \hat{e} \cdot M_{cv}(\textbf{k}) \mid^2}{\epsilon_c(\textbf{k}) - \epsilon_v(\textbf{k})}
\nonumber \\
& & & \times \frac{1}{(\epsilon_c(\textbf{k}) - \epsilon_v(\textbf{k}))^2 - \omega^2}
\end{align}
where $\hat{e}$ is the polarization vector in the direction of the electric field and $M_{cv}(\textbf{k})$ is the dipole matrix element for a transition from a valence state $\epsilon_v(\textbf{k})$ to a conduction state $\epsilon_c(\textbf{k})$.
Eq.~(\ref{eq:epsilon1}) can be obtained starting from the Fermi's golden rule~\cite{Bassani1975}.
Further details are given in the Appendix~\ref{sec:theory}.
However, as can be anticipated from Eq.~(\ref{eq:epsilon1}), the band gap is not a sufficient quantity
to properly describe the data trend and other descriptors have to be included in the analysis.
With this in mind, we map each material onto the simplest system that one can think of for describing
optical transitions: a two-state ($E_1$, $E_2$) system with a transition characterized by (i) an energy $\omega_g=E_2-E_1$, (ii) a probability $K$, and (iii) a degeneracy factor $J=n_1 n_2$, where $n_1$ (resp. $n_2$) is the degeneracy of the state $E_1$ (resp. $E_2$).
In the mapping procedure, which is schematically illustrated in the left panel of Fig.~\ref{fig:mapping}, $\omega_g$ is obtained as the weighted average of the transitions contributing to the optical properties (it will hence be referred to as the average optical gap). $J$ is simply the integral of $j(\omega)$, the corresponding joint density of states (JDOS), and $K$ is the average probability of those transitions.
Their exact analytical expressions are given in the Appendix~\ref{sec:theory}.
In principle, these involve integrals in an extended frequency range.
In practice, one can set an upper frequency limit $\omega_\textrm{max}$ which is high enough compared to the optical absorption processes of interest and which can be identified by considering the optical function $j(\omega)/\omega^3$ (see Eq.~(\ref{omega_max}) of the Appendix~\ref{sec:theory}), as illustrated in the right panel of Fig.~\ref{fig:mapping}.

As a result of the mapping procedure, our data ($n_s$ vs. $E_g^d$) shown in Fig.~\ref{fig:many_models}(b) can be described using the following relationship (see Eq.~(\ref{eq:model_cube}) of the Appendix~\ref{sec:theory}):
\begin{equation}
n_s^2 =  
1 + 8 \pi \frac{K J}{\omega_g^3}  =
1 + \left( \frac{\omega_{\textrm{eff}}}{\omega_g} \right)^3
\label{eq:ns_vs_wg}
\end{equation}
where we have further defined an effective frequency $\omega_{\textrm{eff}}$, which combines $K$ and $J$, in order to ease the analysis.

\begin{figure*}[b!]
\includegraphics{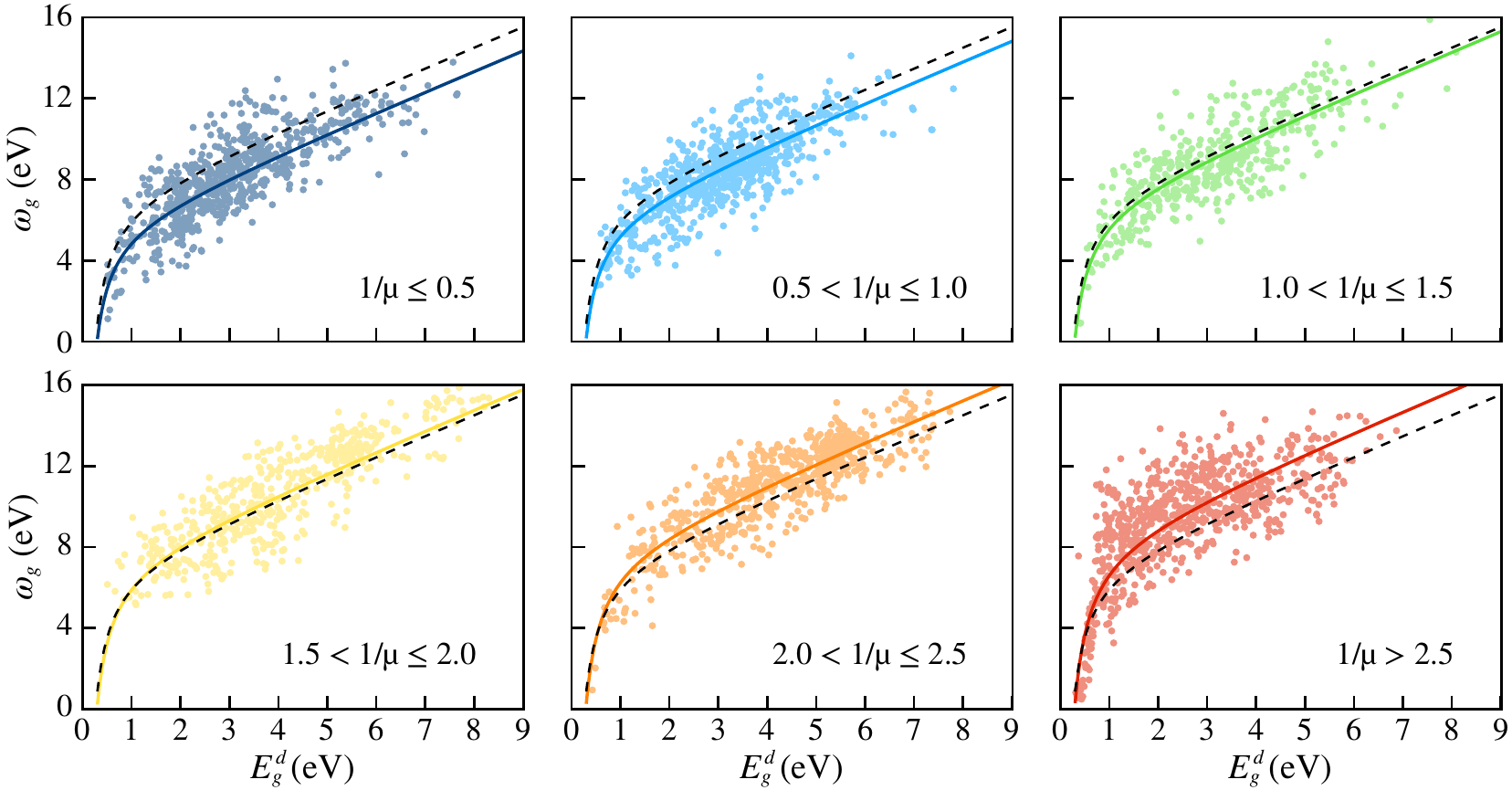}
\caption{
Calculated values of the average optical gap $\omega_g$ as a function of the direct band gap $E_g^d$ (both in eV) for the 4040 materials considered in this study, split considering the different range of the effective mass $1/\mu$.
In each panel, the dashed black line corresponds to $\omega_g = E_g^d + 6.74 - 1.19/E_g^d$ which was obtained by fitting all the data, while the colored line is obtained considering only the data in the subset represented in the panel.}
\label{fig:wg_vs_eg}
\end{figure*}

As can be seen from Fig.~\ref{fig:mapping}, the average optical band gap $\omega_g$ is related to the direct band gap $E_g^d$ by:
\begin{equation}
\omega_g = E_g^d + \Delta.
\label{eq:wg_delta}
\end{equation}
The value of $\Delta$ is material dependent since it is influenced by the dispersion of the valence and conduction bands involved in the transition and their distribution in energy (see Fig.~S3 of the Supplemental Material~\cite{supplemental}) and, indirectly, by the direct band gap $E_g^d$.
The calculated values of $\omega_g$ and $E_g^d$ are shown in Fig.~\ref{fig:wg_vs_eg} for all the materials considered here.

We can describe the relationship between the quantities in Eq.~(\ref{eq:wg_delta}) by the following
equation (see Section~I of the Supplemental Material~\cite{supplemental} for a more detailed description):
\begin{equation}
\omega_g = E_g^d + \alpha \, + \sfrac{\beta}{E_g^d}
\label{eq:wg_vs_eg}
\end{equation}
where $\alpha$=6.74~eV and $\beta$=-1.19~eV$^2$.
However, we note that there is a wide spread of the data around the interpolated value (which translates into a quite large MAE of 1.20~eV for the fit).

This can be traced back to the dependence of $\omega_g$ on the width of the JDOS (see Fig.~S2 of the Supplemental Material~\cite{supplemental}) or, in other words, distribution in energy of the transitions.
The simplest physical quantity that can account for this is the inverse effective mass of the transition~\cite{Fox2001} defined by
\begin{equation}
\frac{1}{\mu} = \frac{1}{m^\ast_v} + \frac{1}{m^\ast_c}
\end{equation}
where $m^\ast_v$ and $m^\ast_c$ are respectively the effective mass of the valence and conduction states averaged over the three possible directions. The details of the calculation of $m^\ast_v$ and $m^\ast_c$ are given in Refs.~\citenum{Hautier2014, Ricci2017}.
By coloring the data points according to 1/$\mu$ in Fig.~\ref{fig:wg_vs_eg}, we note that the larger $\mu$ (the smaller the dispersion of the bands), the smaller $\omega_g$. 
To improve the visualization, the full set of
data has been split according to the values of $1/\mu$.
For each panel, the dashed line represents Eq.~(\ref{eq:wg_vs_eg}) with the the coefficients $\alpha$ and $\beta$ reported above, and the the colored lines represent the same equation by fitting those coefficient considering each subset of data.
The remaining spread in the data (other than the one coming from $\mu$) is difficult to quantify by a simple physical quantity.
Part of it can probably be attributed to the distribution of the bands in energy (see Fig.~S3 of the Supplemental Material~\cite{supplemental}). 

In principle, $\alpha$ and $\beta$ in Eq.~(\ref{eq:wg_vs_eg}) depend on the width of the JDOS.
In practice, in the rest of the paper we assume $\alpha$ and $\beta$ as constants, i.e. considering the fit calculated on the overall set of data ($\alpha$=6.74~eV and $\beta$=-1.19~eV$^2$) to ease the discussion and the analysis.

\begin{figure*}[t!]
\includegraphics{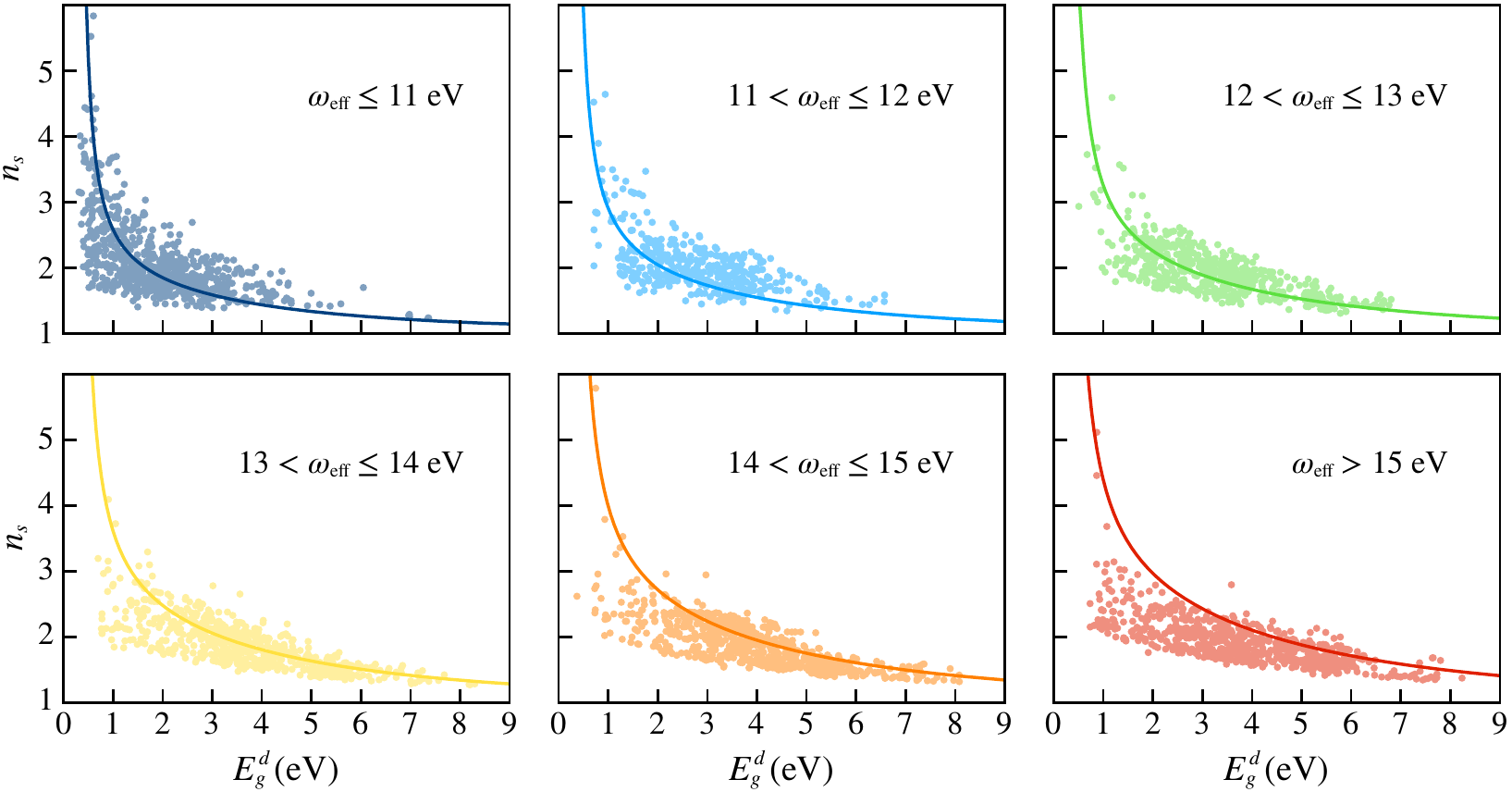}
\caption{
Calculated values of the static refractive index $n_s$ as a function of the direct band gap $E_g^d$ for the 4040 materials considered in this study, split considering the different range of the effective frequency $\omega_\textrm{eff}$.
The solid lines correspond to Eq.~(\ref{eq:ns_vs_eg}) using the the same values of $\omega_{\textrm{eff}}$ reported in Fig.~\ref{fig:many_models}(b).
}
\label{fig:ns_vs_eg}
\end{figure*}

Combining Eqs.~(\ref{eq:ns_vs_wg}) and~(\ref{eq:wg_vs_eg}), we obtain a direct relationship between the average static refractive index $n_s$ and the direct band gap $E_g^d$:
\begin{equation}
n_s = \sqrt{1 + \frac{\omega_{\textrm{eff}}^3}{\left( E_g^d + \alpha \, + \sfrac{\beta}{E_g^d}\right)^3}}
\label{eq:ns_vs_eg}
\end{equation}
which can be compared to all the calculated data, as shown in Fig.~\ref{fig:ns_vs_eg}.
Here, the full set of data has been split according to the values of $\omega_{\textrm{eff}}$ for a better visualization.
In each panel, the colored lines were obtained using the same values of $\omega_{\textrm{eff}}$ indicated in Fig.~\ref{fig:many_models}(b).
Globally, the data follow the trend of Eq.~(\ref{eq:ns_vs_eg}) as represented by these lines, confirming the inverse relationship between refractive index and band gap.
The agreement is quite good given the approximations that are being made for the fit of $\omega_g$ as a function of $E_g^d$.
In particular, the points with large (resp. small) effective masses can fall significantly above (resp. below) the corresponding curve (given that the latter is obtained for an average value of the effective mass).

From our analysis, it is clear that the effective frequency $\omega_{\textrm{eff}}$ (combining the integral of the JDOS $J$ and the average transition probability $K$) and the effective mass $\mu$ (as well as the distribution in energy of the bands) play a key role in counterbalancing the effect of the band gap on the refractive index.
The former is the numerator of the fraction appearing in Eq.~(\ref{eq:ns_vs_eg}), so the larger $\omega_{\textrm{eff}}$ the higher $n_s$.
The latter acts on the denominator by limiting the difference between the direct band gap and the average optical gap: the larger $\mu$, the smaller $\omega_g$ and hence the higher $n_s$.

At this stage, we would like to emphasize that the model that we propose in Eq.~(\ref{eq:ns_vs_eg}) is not predictive.
Indeed, while $\omega_g$ can be determined directly from the electronic structure of the compounds, but $\omega_\textrm{eff}$ (and more precisely $K$) cannot.
We leave it for another study to analyze whether machine learning might help to overcome this limitation.

\subsection{Outliers}

As far as the combination of high refractive index and high band gap is concerned, the most interesting materials are those lying above the curve corresponding to the value  $\omega_{\textrm{eff}}$=12.10~eV, calculated by fitting in the last square sense Eq.~(\ref{eq:ns_vs_eg}) to the full set of our data.
Such materials have either a large value of $\omega_{\textrm{eff}}$ (i.e. following the general trend of the curves) or of $\mu$ (i.e. due to the spread of the data). 
Among those, we found various compounds commonly used for optical devices, a few examples of which are reported in Table~\ref{tab:Table_compounds}. 
In contrast, to the best of our knowledge, some of these outliers have not yet been considered as optical materials (for instance, Ti$_3$PbO$_7$, LiSi$_2$N$_3$, BeS, ...). In Section~VI of the Supplemental Material~\cite{supplemental}, we provide various tables with the 10 materials with the highest refractive index for a given direct band gap range.

Having a high value of the refractive index, the compounds listed in Table~\ref{tab:Table_compounds} also show high response in
the non-linear regime. In particular, LiTaO$_3$, LiNbO$_3$, LiB$_3$O$_5$, and  BaB$_2$O$_4$ are known to have high non-linear
second order coefficients. They are thus commonly used for Second Harmonic Generation (SHG), to convert
the incoming light from UV, or even deep UV, to the visible spectral range (see for example Refs.~\citenum{Boyd2008,Tran2016}).
In contrast, both TiO$_2$ phases (anatase and rutile) are centro-symmetric and they do not show any
response at the second order. But, because of their refractive index, they have been recently investigated as optical
switching devices and waveguides (see for example Refs.~\citenum{Evans2012, Evans2013, Evans2015a}).   

\begin{table*}
\begin{ruledtabular}
\begin{tabular}{r l r r r r r r r}
\multicolumn{1}{c}{Formula} & MP-id & \multicolumn{1}{c}{$n_s$} & \multicolumn{1}{c}{$E_g^d$} & \multicolumn{1}{c}{$\omega_{\textrm{eff}}$} & \multicolumn{1}{c}{$\omega_g$} & \multicolumn{1}{c}{$m^\ast_v$} & \multicolumn{1}{c}{$m^\ast_c$} & \multicolumn{1}{c}{$\mu$} \\
\hline
LiTaO$_3$   & mp-3666   & 2.25 & 3.71 & 14.44 &  9.05 & 3.48 & 1.44 & 1.02\\  
LiNbO$_3$   & mp-3731   & 2.33 & 3.41 & 13.00 &  7.91 & 3.53 & 1.60 & 1.10\\
LiB$_3$O$_5$   & mp-3660   & 1.62 & 6.35 & 16.13 & 13.70 & 6.77 & 1.20 & 1.02\\  
rutile-TiO$_2$   & mp-2657   & 2.84 & 1.78 & 10.88 &  5.66 & 2.53 & 1.00 & 0.72\\  
anatase-TiO$_2$   & mp-390     & 2.60 & 2.35 & 10.73 &  5.96 & 1.95 & 1.85 & 0.95\\  
BaB$_2$O$_4$   & mp-540659 & 1.63 & 4.60 & 13.46 & 11.29 & 15.01 & 0.61 & 0.58\\  
\end{tabular}
\end{ruledtabular}
\caption{
List of known outliers (i.e., lying above the curve corresponding
to $\omega_{\textrm{eff}}$=12.10~eV).
The chemical formula, MP identification (MP-id), average refractive index, direct band gap $E_g^d$ (in eV), the effective frequency $\omega_{\textrm{eff}}$ (in eV), the average optical gap $\omega_g$ (in eV) and the average effective mass of the transitions $\mu$ are shown for each material.
}
\label{tab:Table_compounds}
\end{table*}

\subsection{Trend in oxides}
\label{Results-trend-in-oxides}

We now concentrate on the 3375 oxides. We focus on the chemical composition and the electronic structure of the materials making
the connection with their optical properties.

To properly describe the data distribution, we introduced the effective frequency $\omega_{\textrm{eff}}$ that is related to both
$J$ and $K$ (see Fig.~S5 of the Supplemental Material~\cite{supplemental}). Although both these quantities are important to obtain the correct
$\omega_{\textrm{eff}}$ for each material, only $J$ can be deduced from the electronic structure of the compounds. 
This is the main reason why our model cannot be predictive.
To the best of our knowledge, there is no way to predict the probability transition $K$ just considering the band structure.
Systematic correlations between $K$ and materials properties are still under investigations. 

In order to analyze the trend in terms of their chemistry, the compounds are organized in four different
classes: two groups of transition metal oxides (TMOs), lanthanide oxides, and main-group oxides.
Materials with actinide elements are not taken in account in our analysis.
These classes are created as follows: the groups of the TMOs include compounds in which there is at
least one TM element with the d shell not completely filled and no lanthanide elements. The lanthanide
oxides class contains compounds in which there is at least one lanthanide element, but no TM elements.
Finally, the remaining oxides that do not contain any of the above mentioned elements are included in
the main-group.
The TMOs have been further split in two groups considering not only the TM element but also its
oxidation state~\cite{Waroquiers2017}. In the first group, the d shell of the TM element is empty (e.g. V$^{5+}$) and
therefore the electronic transitions from the top of the
valence to the bottom of the conduction states are expected to be from the O 2p to the TM d-orbitals.
In the second group, the TM d shell is partially filled (e.g. V$^{4+}$, V$^{3+}$) and thus the
transitions are expected to be from the TM filled d states to the empty ones. Finally, all the compounds
that contain TM elements in which the d shell is completely filled (e.g. Zn$^{2+}$, Cu$^+$) are included
in the main group.

For each class, the probability density function is computed for the distribution of the refractive index as function of the band
gap via a Kernel-Density Estimation (KDE) using a Gaussian kernel (see Ref.~\citenum{SCOTT1979} for further details).
The full distribution for each class of materials is reported in Fig.~S6 of the Supplemental Material~\cite{supplemental}.
In Fig.~\ref{fig:ns_vs_eg_classes}, we only represent each class by an ellipse that contains the main data distributions and is obtained as follows.
Its center is located at the average value of the direct gap and refractive index for the corresponding distribution.
The orientation and lengths of its axes are determined using principal component analysis for the materials which belong to the region with a density larger than 75\%.
The curve reported in the figure is obtained from Eq.~(\ref{eq:ns_vs_eg}) using
$\omega_{\textrm{eff}}=12.10~eV$.
As we already stated, materials falling above this curve are the most interesting ones.

The importance of the flatness of the bands at the edges of the valence and conduction bands has already been underlined in Ref.~\citenum{Petousis2017}.
This means that the presence of d and f-orbitals can be helpful.
Indeed, the most interesting compounds (i.e. those located mostly above this curve) come from the first group of TMOs and the
lanthanide oxides, in which those orbitals are present close to the VBM and CBM.
These are the most suitable for applications that require both a wide band gap and a high refractive index.
This is especially true for applications for which the absorption edge is at the limit of the visible region (experimental
$E_g^d$$\sim$3~eV).
For applications in the UV (experimental $E_g^d$$\sim$6~eV), the compounds in the main group of elements reveal to be the most
promising.

In the following subsections, we describe the peculiarities of the four different classes, focusing on a typical example for each of them.
For the four representative materials, we focus on the electronic structure and on a brief description of the optical functions
$j(\omega)$, and $j(\omega)/\omega^3$. We also highlight the different relevant quantities ($E_g^d$, $\omega_g$, and
$\omega_{\textrm{eff}}$).

Before proceeding with the discussion of the different classes, it is worth stressing again that the electronic structures used in this study are taken directly from the Materials Project repository.
They have been obtained in the framework of DFT using the PBE exchange-correlation
functional, that is known to underestimate the band gap with respect to experiments.

\begin{figure}[htbp]
\includegraphics{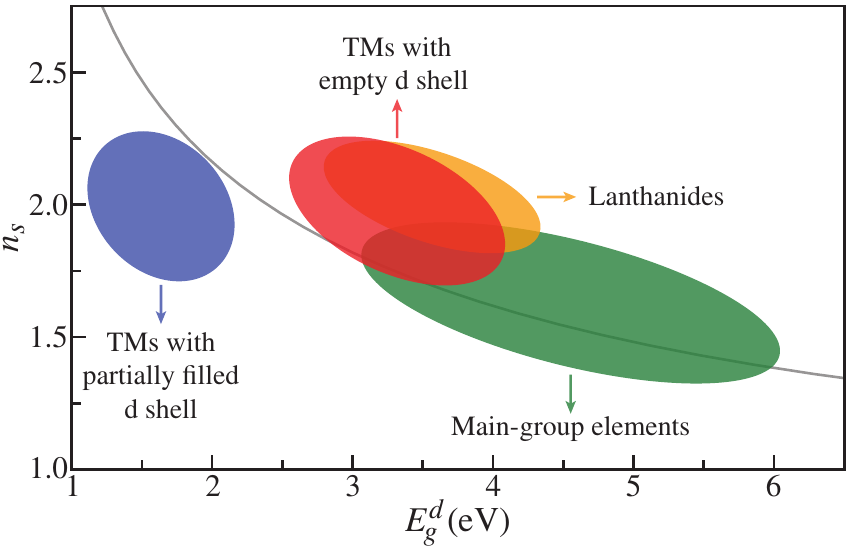}
\caption{
Static refractive index $n_s$ as a function of the direct band gap $E_g^d$ for the four classes of materials (first and second groups of TMOs in red, and blue respectively, main-group elements in green, lanthanides in orange) considered in this study.
Each class is represented by an ellipse (see text) indicating the main distribution of the materials that belong to this class.
The solid line correspond to Eq.~(\ref{eq:ns_vs_eg}) with $\omega_{\textrm{eff}}$=12.10~eV.
}
\label{fig:ns_vs_eg_classes}
\end{figure}

\subsubsection{TMOs with empty d shell (1$^{st}$~group)}
Many materials from this class are of high technological interest as dielectrics and as lenses for optical devices both in the
linear and non-linear regime~\cite{Cox2010}.

\begin{figure*}[h]
\includegraphics{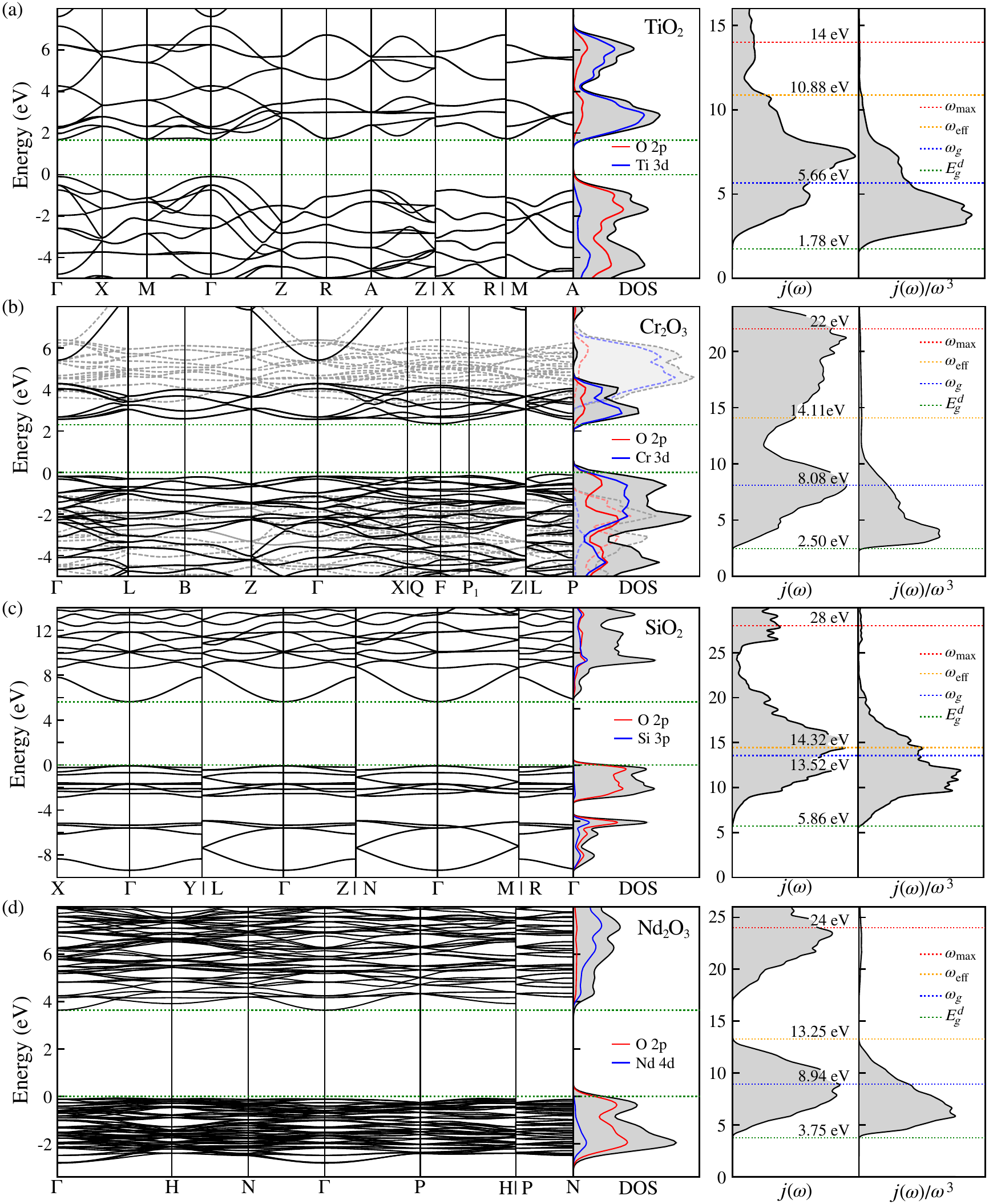}
\vspace{-3mm}
\caption{
Electronic structure [band structure and density of states (DOS)] and optical functions [$j(\omega)$ and $j(\omega)/\omega^3$ in arbitrary units] for (a) TiO$_2$, (b) Cr$_2$O$_3$, (c) SiO$_2$, and (d) Nd$_2$O$_3$.
The direct band gap $E_g^d$, average optical gap $\omega_g$, effective frequency $\omega_{\textrm{eff}}$, and upper limit of integration $\omega_\textrm{max}$ are indicated by green, blue, orange, and red dotted lines.
The four materials have been selected as representatives of the first and second groups of TMOs, the main-group oxides, and lanthanide oxides, respectively.
For Cr$_2$O$_3$ (which shows a magnetic ordering), the electronic structure of both spin components are reported separately (the spin down component is indicated by the use of lighter colors and dashed lines) while the optical functions are the sum of both of them.
}
\label{fig:electronic}
\end{figure*}

As can be seen in Fig.~\ref{fig:ns_vs_eg_classes} (red ellipse), the materials from this class show a relatively high value
for both the refractive index and the band gap.
TiO$_2$ is a typical material of this group. For this compound, the Ti oxidation state is +4 (empty d shell).
This binary oxide compound still generates great interest for the construction of optical devices (see for example
Refs.~\citenum{Evans2015} and~\citenum{Evans2016}).
It is indeed one of the materials with the highest value of the refractive index, while retaining a high transparency throughout
the visible region.
The electronic structure and the optical functions for the rutile phase (mp-2657) are shown in Fig.~\ref{fig:electronic}(a).
These are representative of those for other known TiO$_2$ phases (anatase, brookite, and monoclinic) and for other compounds in this class (e.g. ZrO$_2$, V$_2$O$_5$, LiNbO$_3$, ...).

In all these materials, the bands at the valence and conduction edges are quite flat.
The main contribution to the top valence states originates from the O 2p-orbitals, while that to the bottom conduction bands
comes from the d-orbitals of the transition metal (Ti 3d in the case of TiO$_2$).
The flat nature of the bands at the edge of the band structure in this material can also be appreciated looking at the different
values of the effective masses ($m^\ast_v$=2.53, $m^\ast_c$=1.00, $\mu$=0.72). 
As a consequence the DOS is actually quite high at the band edges.
This leads to an important $j(\omega)$ originating from the transitions from the O 2p-orbitals to the transition metal d
orbitals, and hence to a large value of $J$.
Further, these materials have a wide band gap that arises mainly from electron repulsion effects~\cite{Zaanen1985, Harrison2012}.
This translates into a high refractive index ($n_s$=2.85).

In summary, for transition metals oxides from the first group, the wide band gap (which pushes the refractive index downwards) is
compensated by a large number of available transitions from the top of the valence band to bottom of the conduction bands due to
both the flatness of the band structure and to the high density of states at the band edges.
These materials are thus very interesting candidates for further investigations.

\subsubsection{TMOs with partially filled d shell (2$^{nd}$~group)}
In general, TMOs from the second group have a smaller gap than those from the first group.
This obviously pushes their refractive index upwards.
However, contrary to the first group of TMOs, the majority of the data (blue ellipse in
Fig.~\ref{fig:ns_vs_eg_classes}) fall below the curve.
These materials could be considered good candidates for optical applications that require a moderate
transparency (e.g. in the visible range) and a high refractive index.
We focus on Cr$_2$O$_3$ (mp-19399) as an illustrative example with features common to other compounds of this
class (e.g. PtO$_2$, NiO...).
In this case the Cr oxidation state is +3 (partially filled d shell). 
Chromium oxides are widely used in many sectors such as, for example, catalysis, solar energy
applications, and others (further information can be found in Ref. \citenum{Abdullah2014}).
Since this material shows a magnetic ordering, in Fig.~\ref{fig:electronic}(b) is reported its electronic structure for both spin components separately, and the optical functions resulting from the combination of both of them.
The bands at the edge of the band structure (spin up component) show a flat nature.
This is further emphasized looking at the values of the effective masses ($m^\ast_v$=4.29, $m^\ast_c$=2.53, $\mu$=1.59).
At the bottom of conduction states (spin up component), as in the previous case, the main contribution comes from the d-orbitals of the TM (Cr 3d in the case of Cr$_2$O$_3$).
One of the main difference lies in the contribution of the d-orbitals in the valence states, leading to a more hybridized
character.
The presence of an important amount of d states both at the top of the valence and at the bottom of the conduction leads to a
decrease of the band gap with respect to the TMOs with empty d shell~\cite{Zaanen1985, Harrison2012}. 
Due to the flatness of the bands, the DOS at the edge of the band structure is quite high giving an important $j(\omega)$.
However, in this case, the JDOS gives a more broad spectrum with respect to the TiO$_2$ case, leading to
larger values of both $\omega_g$ and $\omega_{\textrm{eff}}$, and a slightly smaller value of the 
refractive index for Cr$_2$O$_3$ ($n_s$=2.51).

In summary, the main difference with respect to the first group of TMOs lies in the valence bands in which there is a strong
contribution from the d-orbitals.
So, despite their lower gap, the TMOs from the second group show a similar refractive index.

As a final remark, it is worth to mentioning that, far this class of materials, DFT is known to predict wrong band gap and dispersion due to the presence of partially filled d-orbitals. To this purpose, as mentioned in Sec.~\ref{HTMethods}, a Hubbard-like Coulomb $U$ term was added (GGA+$U$) \cite{Dudarev1998, Jain2011a}.

\subsubsection{Main-group oxides}
The main-group oxides (green ellipse in Fig.~\ref{fig:ns_vs_eg_classes}) show a higher diversity than
the other classes. Indeed in this class we can find oxides that contain elements such as Zn, Cd and
Hg in their oxidation state +2 such that they have a full d shell in valence, as well as Si, Ge, etc.
Although the compounds in this class show a common
behavior in terms of their electronic structure, due to the diversity of the materials their band
gaps display a more important spread.

Most of the materials belonging to this class are commonly used as insulators. 
A prototypical example of this class SiO$_2$.
This material is used for many devices and one of the most known applications is found in the amorphous silica phase, used for optical fibers~\cite{Senior2009}.
The electronic structure and the optical functions for the $\beta$-cristobalite $I\bar{4}2d$ tetragonal form (mp-546794) are shown in Fig~\ref{fig:electronic}(d).
Here, as in the case of the first group of transition metals oxides, O 2p states lead to quite flat valence bands and increase the DOS at the valence edge.
In contrast, the conduction bands are very dispersive, showing almost a free-electron like parabolic character which directly translates into the values of the effective masses ($m^\ast_v$=4.15, $m^\ast_c$=0.56, $\mu$=0.49).
This results in a small contribution to the DOS at the bottom of the conduction states.
Consequently, the JDOS $j(\omega)$ [Fig.~\ref{fig:electronic}(c)] does not show any clear peak close to the absorption edge and the refractive index is quite low ($n_s$=1.48). 
Indeed the value of $J$ for this material is smaller than the one of both the representative candidates previously described.
Furthermore, compared to the other cases, the value of $\omega_g$ is much higher than that of $E_g^d$ ($\omega_g>2E_g^d$), and it is much closer to $\omega_{\textrm{eff}}$.

\subsubsection{Lanthanide oxides}
Lanthanide oxides (orange ellipse in Fig.~\ref{fig:ns_vs_eg_classes}) tend to have a wide band gap,
while still showing a high refractive index.
The oxides contained in this class have common features with the TMOs with empty d shell. Indeed, the
two respective ellipses are almost superimposed on each other in Fig.~\ref{fig:ns_vs_eg_classes}. 
As an illustrative example, we have chosen Nd$_2$O$_3$ (mp-1045). These compounds are typically known
because they show good luminescence properties and they can be used as fluorescent materials in lighting
applications (more information can be found in Ref.~\citenum{Bazzi2005161}).
Looking at the electronic structure, the flat nature of the bands can be appreciated at the top of
the valence states coming mainly from the O 2p-orbitals. At the bottom of the conduction states,
there is a slightly more dispersive behavior that comes mainly from the d-orbitals (Nd 4d in the case
of Nd$_2$O$_3$). This is also evident looking at the values of the effective mass
($m^\ast_v$=6.81, $m^\ast_c$=0.53, $\mu$=0.50).
As a result, the DOS at the top of the valence is quite important (as in the case of the 1$^{st}$
group of TMOs), but the slightly more dispersive nature of the bands at the bottom of the conduction
leads to a less pronounced DOS. Anyway, $j(\omega)$ shows a pretty well-defined peak centered at
around $\omega_g$. The large availability of states for an electronic transition from valence to
conduction turns into a large value of $\omega_{\textrm{eff}}$, leading to reasonably high value of the
refractive index ($n_s$=2.06).
All these considerations show the similarity between this class of oxides and the 1$^{st}$ group of
TMOs, and explain why the refractive index remains high in despite the wide band gap.

Finally, we would like to emphasize that the results for the lanthanide oxides need to
be taken with great caution.
Indeed, it is not a simple task to accurately compute materials with f electrons from first
principles relying on pseudopotentials.
In many cases, these electrons are frozen in the core, which may lead to a lack of accuracy.

\section{Conclusions}
\label{Conclusions}
In this study, we have performed a high-throughput investigation of the electronic and optical properties of 4040 semiconductors, calculating their band gap $E_g^d$ and static refractive index $n_s$ in the framework of density functional theory and density functional perturbation theory.
Our data confirm the inverse relationship between $n_s$ and $E_g^d$, but outliers are identified that combine a wide band gap with a high refractive index.
Some of these are well-known optical materials (e.g. TiO$_2$, LiNbO$_3$, ...) while others have never been considered in this framework to the best of our knowledge (e.g. Ti$_3$PbO$_7$, LiSi$_2$N$_3$, BeS, ...).

By mapping all the compounds onto a two-state system, two main descriptors are identified: the average optical gap and the effective frequency.
While the former can be deduced directly from the electronic structure, the latter cannot.
This limits the predictive power of our model and calls for further analysis (e.g. using a machine-learning approach).
However, the model highlights that the decrease of $n_s$ with $E_g^d$ can be partly counterbalanced by a high number and density of available transitions from the top of the valence band to the bottom of the conduction band.
This is directly related to the density of states at the edges of those bands and to the effective mass of such states.

By considering the compounds based on their chemical composition, we have then extracted some common features that can be useful in achieving a wide band gap dielectric.
We have found that materials belonging to the first class of transition metal oxides and lanthanide oxides are the most promising ones for optical applications that require a wide band gap and a high refractive index.

Though our data were collected for materials in the linear regime, they can also be used as a starting point for an analysis of optical properties in the nonlinear regime.
It is worth stressing that our main conclusions are inferred merely from a statistical approach.
Such an approach can help in the understanding and construction of optical devices in a wide range of applications.

\section*{Acknowledgments}

The authors acknowledge X. Gonze and A. Tkatchenko for useful discussions.
FN was funded by the European Union’s Horizon 2020 research and innovation programme under the Marie Sklodowska-Curie grant agreement N$^\circ$ 641640 (EJD-FunMat).
GMR is grateful to the F.R.S.-FNRS for financial support.
GH, GMR and FR acknowledge the F.R.S.-FNRS project HTBaSE (contract N$^\circ$ PDR-T.1071.15) for financial support.
We acknowledge access to various computational resources: the Tier-1 supercomputer of the F\'{e}d\'{e}ration Wallonie-Bruxelles funded by the Walloon Region (grant agreement N$^\circ$ 1117545), and all the facilities provided by the Universit\'{e} catholique de Louvain (CISM/UCL) and by the Consortium des \'{E}quipements de Calcul Intensif en F\'{e}d\'{e}ration Wallonie Bruxelles (C\'{E}CI).

\appendix
\section{Theory}
\label{sec:theory}
\setcounter{figure}{0}
\setcounter{table}{0}
\renewcommand{\thefigure}{A\arabic{figure}}
\renewcommand{\thetable}{A\arabic{table}}

In the linear regime, the dielectric function of a material is the coefficient of proportionality between the macroscopic displacement field $\mathcal{D}$ and the macroscopic electric field $\mathcal{E}$.
In the most general form, both fields are frequency-dependent and they are not necessarily aligned (e.g. in an anisotropic material).
The dielectric function is thus a frequency-dependent tensor $\varepsilon_{\alpha \beta}(\omega)$ (where $\alpha$,$\beta$=1,$\ldots$,3 span the space directions):
\begin{equation}\label{displacement}
\mathcal{D}_{\alpha}(\omega) = 
\sum_{\beta} \varepsilon_{\alpha \beta}(\omega) \mathcal{E}_{\beta}(\omega).
\end{equation}
For sake of simplicity, we will avoid the tensor notation and refer to it simply as $\varepsilon(\omega)$.
In general, the displacement field will include contributions from both electronic and ionic displacements; but in the optical regime, the former dominates. 
This optical (ion-clamped) dielectric permittivity tensor is the one of interest for the present paper.

The refractive index $n(\omega)$ is related to the dielectric function by: 
\begin{equation}\label{ref_index}
n(\omega) = \frac{1}{\sqrt{2}} \sqrt{\varepsilon_1(\omega) +
\sqrt{\varepsilon_1(\omega)^2+\varepsilon_2(\omega)^2}}
\end{equation}
where $\varepsilon_1(\omega)$ and $\varepsilon_2(\omega)$ are the real and imaginary parts of $\varepsilon(\omega)$, respectively.
In the static limit ($\omega=0$), the imaginary part of the dielectric function vanishes for semiconductors, and Eq.~(\ref{ref_index}) becomes:
\begin{equation} \label{static}
n_s=\sqrt{\varepsilon_{1s}}.
\end{equation}

In quantum mechanics, the dielectric function can be related to band-to-band transitions.
Its imaginary part can be obtained from Fermi's golden rule as:
\begin{align}\label{eq:epsilon2}
\varepsilon_2(\omega) = \frac{4\pi^2}{\omega^2} \sum_{v,c}  \int_\mathrm{BZ} & \frac{2d\textbf{k}}{(2\pi)^3}
\mid \hat{e} \cdot M_{cv}(\textbf{k}) \mid^2
\nonumber \\
& \times \delta (\epsilon_c(\textbf{k}) - \epsilon_v(\textbf{k}) - \omega)
\end{align}
where $\hat{e}$ is the polarization vector in the direction of the electric field and $M_{cv}(\textbf{k})$ are the dipole matrix elements for a transition from a valence state $\epsilon_v(\textbf{k})$ to a conduction state $\epsilon_c(\textbf{k})$.
The sum goes over all valence and, in principle, conduction states.
In practice, a convergence test is performed with respect to the number of conduction states to be included in the sum.

The real part of the dielectric function can be derived from its imaginary part via the Kramers-Kronig relations:
\begin{equation}\label{kkrel}
\varepsilon_1(\omega) = 1 + \frac{2}{\pi} \textit{P} \int_{0}^{\infty}
\frac{\omega^{\prime}\varepsilon_2(\omega^{\prime})}{\omega^{\prime2}-\omega^2} \,d\omega^{\prime}
\end{equation}
where \textit{P} indicates the principal part of the integral.
In the static limit, it gives:
\begin{equation}\label{kkrel0}
\varepsilon_{1s} = 1 + \frac{2}{\pi} \int_{0}^{\infty}
\frac{\varepsilon_2(\omega)}{\omega} \,d\omega.
\end{equation}

In principle, the above integral has to be taken from zero to infinity.
In practice, a typical $\varepsilon_2(\omega)$ spectrum usually reveals well-separated peak regions, with little overlap, due to different absorption processes. Therefore, one can set an upper frequency limit $\omega_{max}$ which is high enough compared to the optical absorption processes of interest here, but small compared to other ones.
In this work, $\omega_\textrm{max}$ is defined in such a way that:
\begin{equation} \label{omega_max}
\left(
\int_{0}^{\omega_\textrm{max}}
\frac{\varepsilon_2(\omega)}{\omega} \,d\omega
\middle/
\int_{0}^{\infty}
\frac{\varepsilon_2(\omega)}{\omega} \,d\omega
\right)
\geq 99\%.
\end{equation}

The value of $\varepsilon_{1s}$ and hence $n_s$ can be directly calculated using DFPT at low computational cost~\cite{Baroni1986, Baroni1987, Giannozzi1991, Gonze1997, Gonze1997a}.
Indeed, conduction states do not need to be taken into account in contrast with the sum over states formulation within the random-phase approximation~\cite{Adler1962,Wiser1963}.
The drawback of the DFPT approach is that only the static limit of the dielectric function is computed and hence the frequency dependence is not available.
This can be partly circumvented as follows.

\begin{figure}[h]
\includegraphics{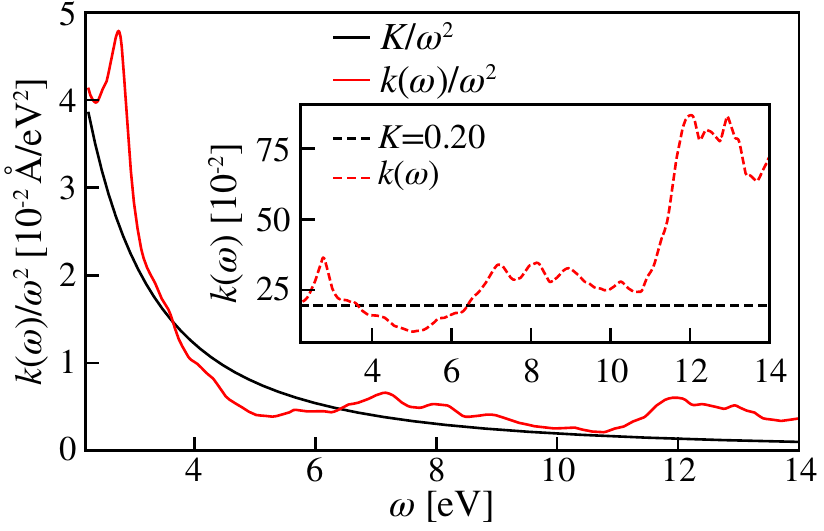}
\caption{Comparison of the frequency-dependent
transition probability $k(\omega)/\omega^2$ and its constant value $K/\omega^2$ obtained from Eq.~(\ref{eq:k_constant}) for a real material (TiO$_2$, mp-2657). In the inset the comparison
is given for $k(\omega)$ and K.  All functions are shown in a frequency range [2, $\omega_{max}$], with $\omega_{max}$ equal to 14 eV for this material.
}
\label{fig:transition_probability}
\end{figure}

We first introduce the joint density of states (JDOS):
\begin{equation}
 j(\omega) = \sum_{v,c} \int_\mathrm{BZ} \frac{2d\textbf{k}}{(2\pi)^3}
 \delta (\epsilon_c(\textbf{k}) - \epsilon_v(\textbf{k}) - \omega),
\end{equation}
which can easily be obtained from DFT calculations of the electronic band structure.
We note its similarity with Eq.~(\ref{eq:epsilon2}).
As a result, we define a frequency-dependent transition probability $k(\omega)$ such that:
\begin{equation}
\varepsilon_2(\omega)=\frac{4\pi^2}{\omega^2} k(\omega) j(\omega).
\end{equation}
We note that, if the matrix elements $\mid \hat{e} \cdot M_{cv}(\textbf{k}) \mid^2$ were all equal to a constant $K$, we would simply have $k(\omega)=K$.
In Fig.~\ref{fig:transition_probability}, we show, as an example, a comparison between the frequency-dependent transition probability $k(\omega)$ and the constant value $K$ for a real material.

\begin{figure}[h]
\includegraphics{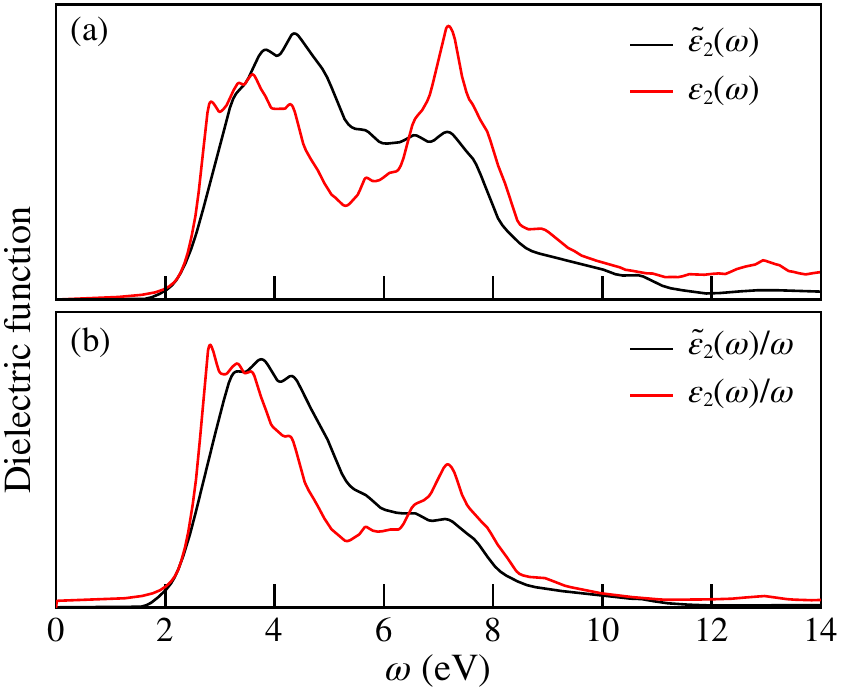}
\caption{Comparison of the imaginary part of (a) the dielectric function
$\varepsilon_2(\omega)$ and (b) $\varepsilon_2(\omega)/\omega$ considering the two methodologies
of calculation for a real material (TiO$_2$, mp-2657). The red curves are obtained averaging
the diagonal components of the DFT
imaginary part of the dielectric function (Eq.~(\ref{eq:epsilon2})). The black curves are obtained via a renormalization of the 
$j(\omega)$ (Eq.~(\ref{eq:eps_to_jdos})).}
\label{fig:diel_func}
\end{figure}

Consequently, a simple approximation for the imaginary part of the dielectric function can be obtained as~\cite{Bassani1975}:
\begin{equation}
\label{eq:eps_to_jdos}
\tilde{\varepsilon}_2(\omega)=4\pi^2 K\frac{j(\omega)}{\omega^2}.
\end{equation}
The value of $K$ is determined such that $\tilde{\varepsilon}_2(\omega)$ also satisfies the Kramers-Kronig relation given by Eq.~(\ref{kkrel0}).
This is strictly equivalent to defining $K$ as a weighted average of $k(\omega)$ as follows:
\begin{equation}
K=\left.
\int_{0}^{\omega_\textrm{max}} k(\omega) \frac{j(\omega)}{\omega^3} d\omega
\middle/
\int_{0}^{\omega_\textrm{max}} \frac{j(\omega)}{\omega^3} d\omega
\right.
\label{eq:k_constant}
\end{equation}
A comparison of $\varepsilon_2(\omega)$ with $\tilde{\varepsilon}_2(\omega)$ is given in
Fig.~\ref{fig:diel_func}.

Using this approximation for the imaginary part of the dielectric function, Eq.~(\ref{kkrel0}) can be rewritten as follows:
\begin{align}\label{jintegral}
\varepsilon_{1s} & = 1 + \frac{2}{\pi}
\int_{0}^{\omega_\textrm{max}} \frac{\tilde{\varepsilon}_2(\omega)}{\omega^2} \,d\omega
\nonumber \\
& = 1 + 8 \pi K
\int_{0}^{\omega_\textrm{max}} \frac{j(\omega)}{\omega^3} \,d\omega.
\end{align}

Introducing the integral of the JDOS $J$, we further define the effective frequency $\omega_\textrm{eff}$:
\begin{align}
\omega_\textrm{eff} & = 
\left(  \frac{2}{\pi} \int_{0}^{\omega_\textrm{max}} \omega^2 \tilde{\varepsilon}_2(\omega) \,d\omega \right)^{\frac{1}{3}} \nonumber \\
& = 
 \left( 8 \pi K \int_{0}^{\omega_\textrm{max}} j(\omega) \,d\omega\right)^{\frac{1}{3}} = \left( 8 \pi K J \right)^{\frac{1}{3}}
\label{eq:omegaeff}
\end{align}
and the average optical gap $\omega_g$:
\begin{align}
\omega_g & =\left(
\int_{0}^{\omega_\textrm{max}}
\omega^2 \tilde{\varepsilon}_2(\omega) \,d\omega
\middle/
\int_{0}^{\omega_\textrm{max}}
\frac{\tilde{\varepsilon}_2(\omega)}{\omega} \,d\omega
\right)^{\frac{1}{3}} \nonumber \\
& = \left(
\int_{0}^{\omega_\textrm{max}}
 j(\omega) \,d\omega
\middle/
\int_{0}^{\omega_\textrm{max}}
\frac{j(\omega)}{\omega^3} \,d\omega
\right)^{\frac{1}{3}}.
\label{eq:wg}
\end{align}
Finally, we can thus write:
\begin{equation}
n_s^2 = \varepsilon_{1s} = 1 + \left(\frac{\omega_\textrm{eff}}{\omega_g}\right)^3,
\label{eq:model_cube}
\end{equation}
which is Eq.~(2) in the main text.

\bibliography{HT_lineardb}
\bibliographystyle{apsrev4-1}

\end{document}